\documentclass[twocolumn]{aastex6}
\defcitealias{keane2016}{K16}
\slugcomment{Accepted to ApJL}
\shorttitle{Interstellar Scintillation and the Radio Counterpart of the Fast Radio Burst FRB150418}
\shortauthors{K.~Akiyama \& M.~D.~Johnson}
\newcommand{\haystack}{1}
\newcommand{\naoj}{2}
\newcommand{\cfa}{3}
\newcommand{\jsps}{4}
\begin{document}
\title{Interstellar Scintillation and the Radio Counterpart of the Fast Radio Burst FRB150418}
%
\author{Kazunori Akiyama\altaffilmark{\haystack,\naoj,4}}
\author{Michael D. Johnson\altaffilmark{\cfa}}
%
\altaffiltext{\haystack}{Massachusetts Institute of Technology, Haystack Observatory, Route 40, Westford, MA 01886, USA}
\altaffiltext{\naoj}{Mizusawa VLBI Observatory, National Astronomical Observatory of Japan, 2-21-1 Osawa, Mitaka, Tokyo 181-8588, Japan}
\altaffiltext{\cfa}{Harvard Smithsonian Center for Astrophysics, 60 Garden Street, Cambridge, MA 02138, USA}
\altaffiltext{\jsps}{\url{kazu@haystack.mit.edu}; JSPS Postdoctoral Fellow for Research Abroad}
%
\begin{abstract}
	\citet{keane2016} have recently reported the discovery of a new fast radio burst, FRB150418, with a promising radio counterpart at 5.5 and 7.5~GHz -- a rapidly decaying source, falling from 200-300~$\mu$Jy to 100~$\mu$Jy  on timescales of $\sim$6~d. This transient source may be associated with an elliptical galaxy at redshift $z=0.492$, providing the first firm spectroscopic redshift for a FRB and the ability to estimate the density of baryons in the intergalactic medium via the combination of known redshift and radio dispersion of the FRB. 
	An alternative explanation, first suggested by \citet{williams2016}, is that the identified counterpart may instead be a compact AGN. The putative counterpart's variation may then instead be extrinsic, caused by refractive scintillation in the ionized interstellar medium of the Milky Way, which would invalidate the association with FRB150418. We examine this latter explanation in detail and show that the reported observations are consistent with scintillating radio emission from the core of a radio-loud active galactic nucleus (AGN) having a brightness temperature $T_{\rm b} \gtrsim 10^9~{\rm K}$. 
	Using numerical simulations of the expected scattering for the line of sight to FRB150418, we provide example images and light curves of such an AGN at 5.5 and 7.5~GHz. These results can be compared with continued radio monitoring to conclusively determine the importance of scintillation for the observed radio variability, and they show that scintillation is a critical consideration for continued searches for FRB counterparts at radio wavelengths. 
\end{abstract}
%
%
\section{Introduction\label{sec:introduction}}

Fast radio bursts (FRBs) are highly-dispersed radio transients emitting a radio flux density of ${\sim}1$~Jy for only a few milliseconds or less. Since the first discovery reported in 2007 \citep{lorimer2007}, more than 17 FRBs\footnote{\url{http://www.astronomy.swin.edu.au/pulsar/frbcat/}} \citep[][]{petroff2016} have been discovered with the Parkes Radio telescope \citep[e.g.][]{lorimer2007,keane2012,keane2016}, the Arecibo observatory \citep{spitler2014,spitler2016}, and the Green Bank telescope \citep{masui2015}. This population of FRBs is highly inhomogeneous, with some showing high circular polarization \citep{petroff2015}, others showing high linear polarization \citep{masui2015}, and others with little polarization \citep{keane2016}. Although most detected FRBs are isolated events, one source has recently been found to be repeating \citep{spitler2016}.  

The origin of the FRBs remains uncertain. Their large dispersion measures, ${\sim} 400-1600$~cm$^{-3}$\,pc, exceed expected values for propagation in the interstellar medium, suggesting that FRBs are extragalactic at redshifts of $z\sim 0.5-1$. An extragalactic pulsed signal enables a direct probe of properties of the intergalactic medium (IGM) through the cold plasma dispersion relationship, just as pulsars are used to probe the ionized interstellar medium (ISM) of the Milky Way. The short durations and extreme brightness temperatures of FRBs suggest compact progenitors, such as the magnetars \citep[e.g.][]{lyubarsky2014}, neutron stars \citep[e.g.][]{totani2013,falcke2014,zhang2014}, white-dwarfs \citep[e.g.][]{kashiyama2013}, and black holes \citep[e.g.][]{keane2012}, although the repeating FRB is incompatible with the many proposed cataclysmic events.  

Recently, \citet{keane2016} (hereafter \citetalias{keane2016}) reported the discovery of a new FRB (FRB150418) followed by a slower radio transient detected with the Australia Telescope Compact Array (ATCA). The radio light curves of this transient (hereafter J0716-1900) at 5.5 and 7.5~GHz show rapid decay on timescales of $\sim$6~d. Optical observations with {\it Subaru} found that J0716-1900 is associated with the elliptical galaxy WISE J071634.59-190039.2 at the redshift of 0.492$\pm$0.008. If this galaxy is indeed associated with FRB150418, then its identification provides the first measured redshift of a FRB. The dispersion measure of the FRB then provides an estimate of the density of the IGM, in this case giving a value consistent with the standard $\Lambda$CDM cosmology.

However, scintillation in the ionized interstellar medium can also introduce rapid variability of compact radio sources such as active galactic nuclei (AGNs), pulsars and masers \citep[see; e.g.][]{RCB_1984,rickett1990,narayan1992,lovell2008}. Indeed, follow-up observations of J0716-1900 by \citet{williams2016} detected rapidly varying flux density, with some measurements as high as the original detections by \citetalias{keane2016} following the FRB. Moreover, because the line of sight to FRB150418 lies close to the Galactic plane (Galactic latitude $b\approx -3.2 ^\circ$), it has scattering that is significantly stronger than that at larger latitudes. Consequently, as noted by \citetalias{keane2016}, the effects of scintillation are significant for this FRB and other nearby compact sources and must be considered as a potential source of the rapid variation seen in J0716-1900. 
If the variation is indeed extrinsic, caused by scintillating emission from a compact galactic nucleus, then its association with FRB150418 is likely spurious. 

In this Letter, we study the role of interstellar scintillation in the radio variability of J0716-1900. In \S\ref{sec:2}, we briefly introduce the theory of the interstellar scattering and give expected scattering properties along the line of sight to FRB150418. Then, we present theoretical simulations based on these quantities in \S\ref{sec:3}. Finally, we discuss the role of continued follow-up observations of FRB150418 and general considerations for follow-up of future FRBs in \S\ref{sec:4}. Throughout the Letter, we use a $\Lambda$CDM cosmology with $h=0.705$, $\Omega _m=0.228$ and $\Omega _\Lambda = 0.726$ \citep{komatsu2009}, providing the luminosity distance and linear scale of 2808~Mpc and 6.1~pc mas$^{-1}$ for J0716-1900, respectively.

\section{Scattering and Scintillation of J0716-1900}\label{sec:2}
\subsection{Scattering Theory\label{subsec:theory}}
Scattering of radio waves in the interstellar plasma arises from fluctuations in electron density. A variety of measurements find that the three-dimensional power spectrum $P(\mathbf{q})$ of the density fluctuations corresponds to a turbulent cascade that is injected on scales of ${\gg}1$~AU and is dissipated on scales of ${\sim}10^2$~km: $P(\mathbf{q}) \propto |\mathbf{q}|^{-(\alpha+2)}$, with $\alpha$ close to the value of $5/3$ expected for Kolmogorov turbulence \citep[][]{armstrong1995}. In many instances, the scattering properties can be well described by a single, thin phase-changing screen $\varphi ({\bf x})$, where ${\bf x}$ is a transverse coordinate on the screen. The statistical characteristics of the scattering and scintillation can then be related to statistical characteristics of the phase fluctuations through a spatial structure function $D_\varphi ({\bf x})=\langle[\varphi ({\bf x}+{\bf x_0})-\varphi ({\bf x_0})]^2\rangle_{{\bf x_0}}$. Density fluctuations that follow the above power-law then give rise to a power-law structure function, $D_\varphi ({\bf x})\propto |{\bf x}|^\alpha$.

The properties of the scattering screen are characterized by a pair of length scales. The \emph{phase coherence length} (or diffractive scale), $r_0\propto\lambda^{-2/\alpha}$, decreases with increasing observing wavelength $\lambda$ and determines the scale at which the screen phase decorrelates: $D_\varphi (r_0)\equiv 1$. The \emph{Fresnel scale}, $r_{\rm F}\equiv \sqrt{\frac{\lambda D}{2\pi}}$, depends on the distance $D$ from the observer to the scattering screen and determines how the geometrical phase of propagation varies across the screen. For radio observations, interstellar scattering is usually in the strong scattering regime, corresponding to the condition $r_0 \ll r_{\rm F}$ (i.e., $D_\varphi (r_{\rm F}) \gg 1$), and a third scale becomes important: the \emph{refractive scale}, $r_{\rm R}=r_{\rm F}^2/r_0\propto \lambda ^{1+2/\alpha}$, which determines the size of the scattered image of the point source.

Scintillation in strong scattering is dominated by two distinct branches, \emph{diffractive} and \emph{refractive}, on these widely separated scales. Diffractive scintillation, arising from fluctuations on scales of $r_0$, is quenched by a source exceeding the angular scale $r_0/D$. As a result, diffractive  scintillation is typically quenched by AGN. Refractive scintillation arises from fluctuations on scales of $r_{\rm R}$ and is only quenched by a source exceeding the angular scale $r_{\rm R}/D$ so can persist for compact AGN. With the characteristic transverse velocity of the scattering material $v_\perp$, the \emph{diffractive}  and \emph{refractive timescales} are given by $t_0\equiv r_0/v_\perp$ and $t_{\rm R}\equiv r_{\rm R}/v_\perp$. For pulsars, $v_\perp \sim 10^7$~cm~s$^{-1}$ is typical \citep[e.g.][]{cordes1988,rickett1990}, but this is typically dominated by proper motion of the pulsar. For an extragalactic source, the velocity is determined by a combination of motion of the Earth and of the scattering material, and we adopt a characteristic velocity of $v_\perp=5\times 10^6$~cm~s$^{-1}$ \citep[see, e.g.,][]{Rickett_1995}.

The flux variability due to the scattering effects is often quantified with the \emph{modulation index}, defined by $m\equiv \sqrt{\langle F_\nu ^2\rangle-\langle F_\nu \rangle ^2}/\langle F_{\nu}\rangle$, where $F_\nu$ is the flux density. For refractive scintillation, the modulation index for a source with angular size $\theta_{\rm src}$ smaller than $\theta_{\rm scatt}$ (i.e., the refractive scale) is $m \approx (r_0/r_{\rm F})^{2-\alpha}$, where the precise prefactor is of order unity \citep[e.g.][]{Goodman_1985,narayan1992}. Larger sources suppress the modulation index by a factor of $(\theta _{\rm scatt}/\theta _{\rm src})^{2-\alpha/2}$, where $\theta_{\rm src}$ is the unscattered source size. 

\subsection{Expected Scattering Properties for J0716-1900}\label{subsec:prm}

\begin{table}[t]
	\centering{}
	\caption{Estimated scattering properties for J0716-1900
		\label{tab:prm}}
	\begin{tabular}{clccc}
		\hline \hline
		Quantity & Unit & 1 GHz & 5.5~GHz & 7.5~GHz \\
		\hline
		$\theta _{\rm scatt}$ & (mas) & 4.5 & 0.11 & 0.054\\ 
		$r_0$ & (cm) & $2.2\times 10^{8}$  & $1.7\times 10^{9}$ & $2.4\times 10^{9}$\\
		$r_{\rm F}$ & (cm) & $1.2\times 10^{11}$ & $5.2\times10^{10}$ & $4.4\times10^{10}$\\
		$r_{\rm R}$ & (cm) & $6.8\times 10^{13}$ & $1.6\times10^{12}$ & $8.0\times10^{11}$\\
		$t_0$ & (s) &  44 & 340 & 490\\
		$t_{\rm R}$ & (d) & 160 & 3.7 & 1.9\\
		$m$ & (\%) &  12 & 32 & 38\\
		\hline
		\vspace{1em}
	\end{tabular}
\end{table}

We can estimate characteristic scattering properties for J0716-1900 with the Galactic free electron density model, NE2001\footnote{\url{http://www.nrl.navy.mil/rsd/RORF/ne2001/}} \citep{cordes2002,cordes2003}. For this model, the expected FWMH scattering size along the line of sight to J0716-1900 at 1~GHz is $\theta_{\rm scatt,1{\rm GHz}}=4.5$~mas. Due to its low Galactic latitude, this value is significantly higher than the median angular broadening at 1~GHz (${\sim}1~{\rm mas}$). Leaving $\theta_{\rm scatt,1{\rm GHz}}$ as a free parameter to provide formulas that are applicable to arbitrary lines of sight (but referencing to our assumed value for J0716-1900), other scattering parameters can then be estimated as follows:
\begin{eqnarray*}
	\ \theta_{{\rm scatt}} & = & 4.5\,{\rm mas}\times\left(\frac{\theta_{{\rm scatt},1{\rm GHz}}}{4.5\,{\rm mas}}\right)\left(\frac{\nu}{1\,{\rm GHz}}\right)^{-1-\frac{2}{\alpha}}\\
	r_{0} & = & 2.1\times10^{8}\,{\rm cm}\times
	\left(\frac{\theta_{{\rm scatt},1{\rm GHz}}}{4.5\,{\rm mas}}\right)^{-1}
	\left(\frac{\nu}{1\,{\rm GHz}}\right)^{\frac{2}{\alpha}}\\
	r_{{\rm F}} & = & 1.2\times10^{11}\,{\rm cm}\times
	\left(\frac{\nu}{1\,{\rm GHz}}\right)^{-\frac{1}{2}}
	\left(\frac{D}{1\,{\rm kpc}}\right)^{\frac{1}{2}}\\
	r_{\rm R} & = & 6.8\times10^{13}\,{\rm cm}\times
	\left(\frac{\theta_{{\rm scatt},1{\rm GHz}}}{4.5\,{\rm mas}}\right)
	\left(\frac{\nu}{1\,{\rm GHz}}\right)^{-1-\frac{2}{\alpha}}\\
	&&\hspace{6em}\times
	\left(\frac{D}{1\,{\rm kpc}}\right)\\
	t_{0} & = & 44\,{\rm s}\times
	\left(\frac{\theta_{{\rm scatt},1{\rm GHz}}}{4.5\,{\rm mas}}\right)^{-1}
	\left(\frac{\nu}{1\,{\rm GHz}}\right)^{\frac{2}{\alpha}}
	\left(\frac{v_{\perp}}{50\,{\rm km\,s^{-1}}}\right)^{-1}\\
	t_{\rm R} & = & 160\,{\rm d}\times
	\left(\frac{\theta_{{\rm scatt},1{\rm GHz}}}{4.5\,{\rm mas}}\right)
	\left(\frac{\nu}{1\,{\rm GHz}}\right)^{-1-\frac{2}{\alpha}}
	\left(\frac{D}{1\,{\rm kpc}}\right)\\
	&&\hspace{2.5em}\times
	\left(\frac{v_{\perp}}{50\,{\rm km\,s^{-1}}}\right)^{-1}\\
	m & \approx & 12\,\%\times
	\left(\frac{\theta_{{\rm scatt},1{\rm GHz}}}{4.5\,{\rm mas}}\right)^{-(2-\alpha)}
	\left(\frac{\nu}{1\,{\rm GHz}}\right)^{\frac{4}{\alpha} - \frac{\alpha}{2} - 1}\\
	&&\hspace{2.2em}\times
	\left(\frac{D}{1\,{\rm kpc}}\right)^{-\frac{1}{6}}.
\end{eqnarray*}
In Table \ref{tab:prm}, we give characteristic quantities for a screen at $D=1$~kpc at 1, 5.5 and 7.5~GHz with a power-law index of $\alpha = 5/3$. We emphasize that these results are only appropriate in the strong-scattering regime, so for J0716-1900 they are applicable at frequencies below ${\sim}40\,{\rm GHz}$. This weak/strong transition frequency is much higher than that of most lines of sight because of the low Galactic latitude of J0716-1900.

The most important implication from Table \ref{tab:prm} is that J0716-1900 can be highly affected by refractive scattering at 5.5 and 7.5~GHz, if the source is more compact than ${\sim} 0.1$~mas (corresponding to a distance of $\sim 0.6$~pc at the redshift of $z=0.492$).  This upper-limit is reasonable for the radio core of a relativistic jet \citep[e.g., as has been seen directly with space VLBI;][]{horiuchi2004}. We also note that recent VLBI surveys have shown that  fainter sources are more likely to be dominated by compact components \citep[e.g.][]{deller2014}, and preliminary results of VLBI observations show that J0716-1900 is unresolved on milliarcsecond scales \citep{bassa2016,marcote2016}. 
Refractive scattering would then cause modulation of ${\sim}30\%$ on a timescale of a few days, similar to what has been observed (see Figure~\ref{fig:lc1}). \citet{williams2016} have also obtained a similar conclusion. 

Thus, in addition to the importance of scintillation for the radio variability a compact afterglow, as noted by \citetalias{keane2016}, refractive scintillation is also a critical consideration even for compact emission from an AGN, and the radio variability of J0716-1900 is comparable to the expected refractive scintillation in the ISM. Note also that the pulse broadening due to the interstellar scintillation is only $\sim$0.02~ms at 1~GHz, which is much shorter than the observed pulse duration of $0.8\pm0.3$~ms for FRB150418 \citepalias{keane2016}.

\section{Numerical Simulations}\label{sec:3}
\begin{figure*}
	\centering{}
	\includegraphics[width=0.9\textwidth]{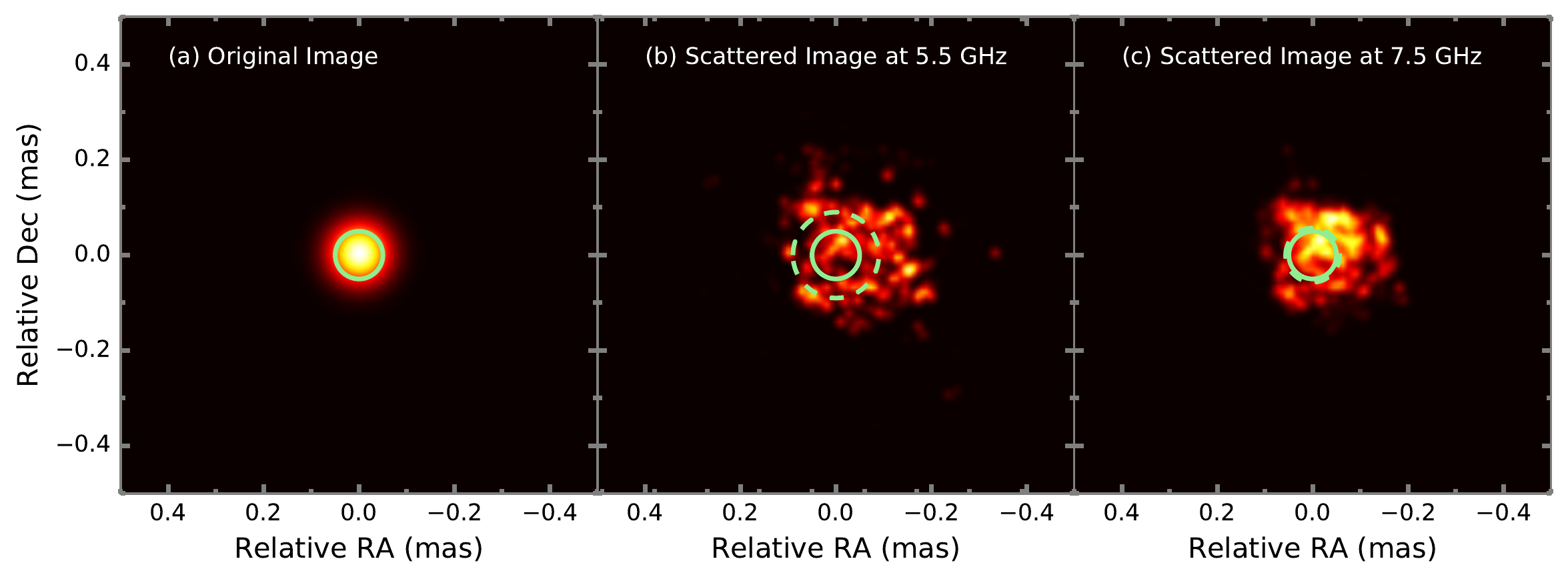}
	\caption{
		Simulated images showing the effects of refractive substructure at $\nu = 5.5$ and $7.5$~GHz. Panel (a) shows the circular Gaussian intrinsic source with the FWHM size of 0.1~mas, which is shown in a circle with the solid line. Panels (b) and (c) show snapshot images of the scattered structure at 5.5 and 7.5~GHz. The scattering parameters correspond to the NE2001 estimates for J0716-1900 (see \S\ref{subsec:prm}). The dashed lines indicate the ensemble-average scattered size, $\sqrt{\theta_{\rm scatt}^2 + \theta_{\rm src}^2}$. Other effects from refractive scattering are also apparent, such as shifts in the image centroids.
		\label{fig:images}}
\end{figure*}

To further study the radio-flux variation caused by the interstellar scintillation and understand how the variations may be correlated at 5.5 and 7.5~GHz, we performed numerical simulations of the scattering of J0716-1900. Following \citet{johnson2015}, we generated scattered images at 5.5 and 7.5~GHz for an intrinsic source that was a circular Gaussian with a full-width-at-half-maximum (FWHM) size of 0.1~mas.\footnote{Because the scattering has deterministic frequency dependence, a single scattering screen determines the scattered image at all frequencies.} We generated a scattering screen with $2^{14}\times 2^{14}$ random phases with the characteristic scattering parameters given in Table~\ref{tab:prm}. Our simulations span 1207~d ($\sim$ 300-600~$t_R$) with a time resolution of 0.27~d ($\ll$ $t_R$). To investigate the variability statistics, we also generated $\sim 200$ different scattering realizations, providing a total span of $\sim 2\times10^5$ d. We show example images of the scattered structure for one realization of the scattering in Fig.~\ref{fig:images}. We calculated the total flux density of each scattered image at each time to generate a light curve for each frequency.

\begin{figure*}
	\centering{}
	\includegraphics[width=0.9\textwidth]{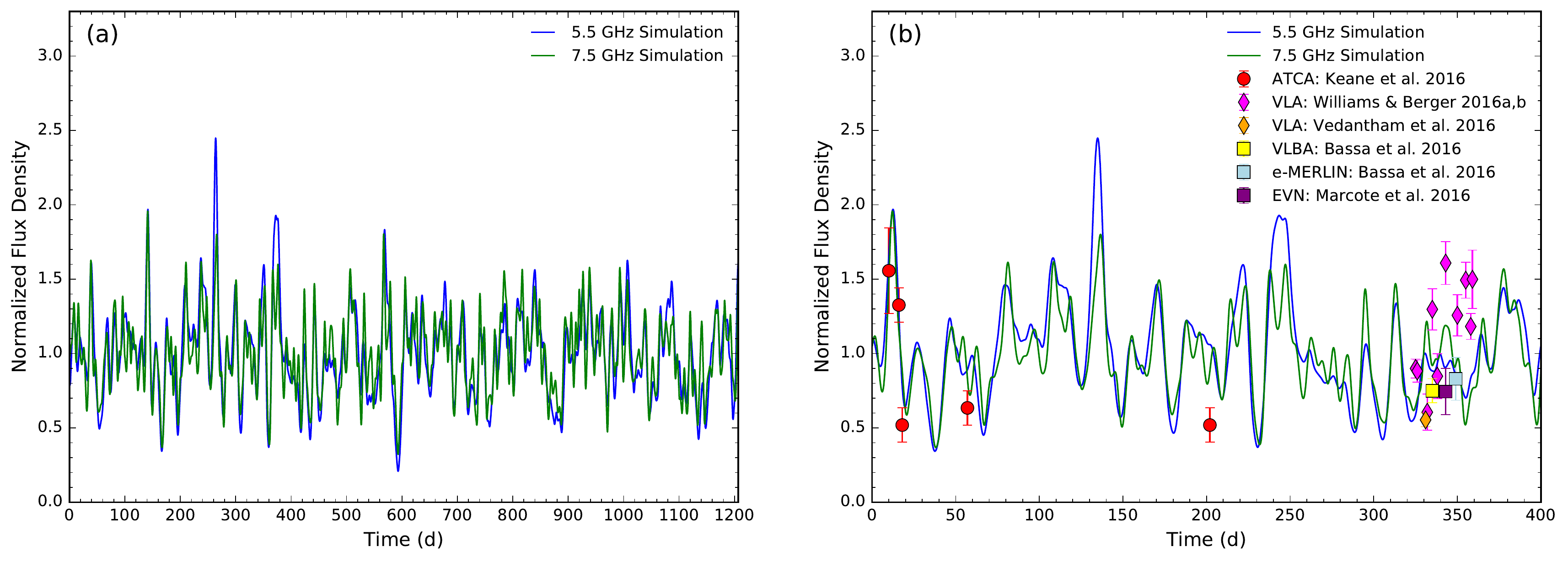}
	\caption{
		Simulated light curves at 5.5~GHz (blue line) and 7.5~GHz (green line) for 1207~d (a) and for 400~d following a flare-like scintillation feature (b). Each light curve is normalized by its mean value. As expected, the fluctuations at these two frequencies are tightly correlated.  
		For reference, the 5.5~GHz data of \citetalias{keane2016} with ATCA, \citet{williams2016,williams2016b} and \citet{Vedantham_2016} with VLA, \citet{bassa2016} with VLBA and e-MERLIN, and \citet{marcote2016} with EVN are also shown after being normalized by the mean value of all data.
		\label{fig:lc1}}
\end{figure*}

\begin{figure*}
	\centering{}
	\includegraphics[width=0.9\textwidth]{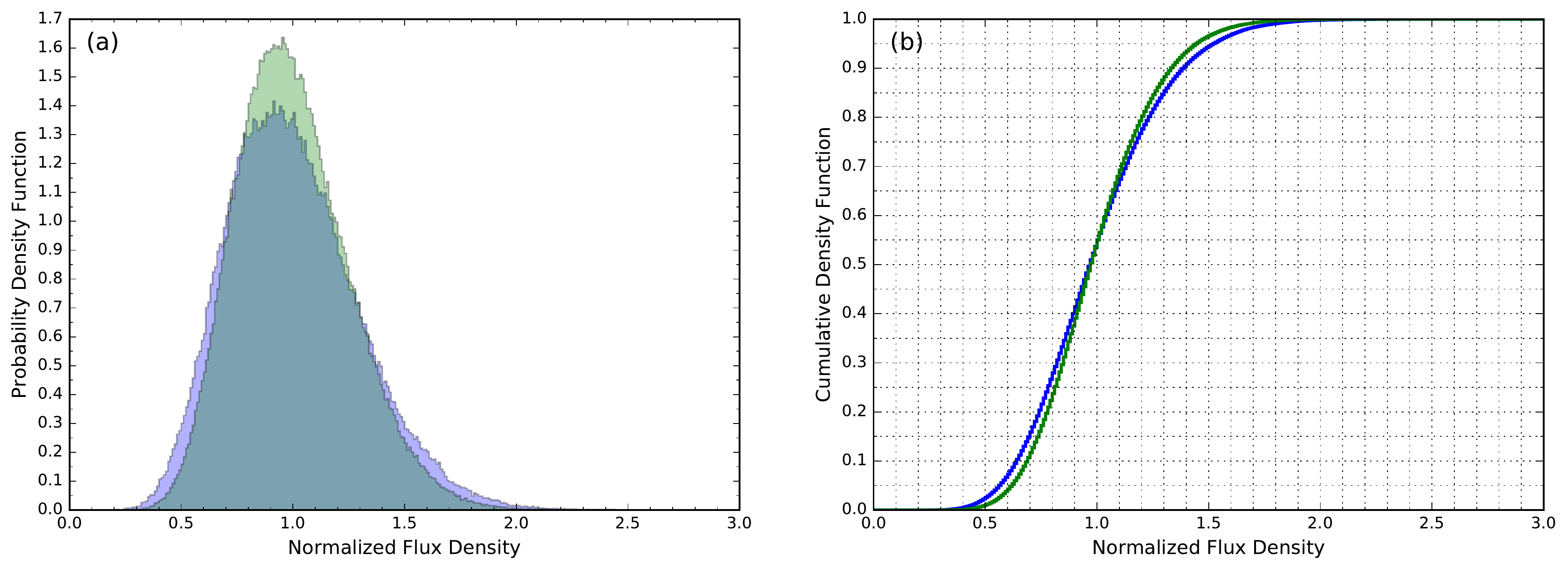}
	\caption{
		The probability distributions of the radio flux at 5.5~GHz (blue line) and 7.5~GHz (green line) obtained from all simulated data (a total of ${\sim}2\times10^5$ d) and normalized by their mean values. Panel (a) shows the probability density function (PDF); panel (b) shows the cumulative distribution function (CDF). For a time-independent intrinsic source, the PDF of a complete scattering ensemble is expected to follow a Rice distribution. 
		\label{fig:lc2}}
\end{figure*}

In Figure~\ref{fig:lc1}, we show simulated light curves at 5.5 and 7.5~GHz, and in Figure~\ref{fig:lc2} we show the probability distribution for the simulated light curves. The light curves exhibit the expected fast variability discussed in \S\ref{subsec:prm}. The modulation index is ${\sim} 29\%$ and ${\sim} 25\%$ for the whole data at 5.5 and 7.5~GHz, respectively, consistent with the original observations of the ATCA \citepalias{keane2016} and also with the follow-up observations with the Very Large Array (VLA) \citep{williams2016b,williams2016,Vedantham_2016}. For direct comparison with these observations, we also compare light curves for 400 d with the normalized light curve of \citetalias{keane2016}, \citet{williams2016,williams2016b}, \citet{Vedantham_2016}, \citet{bassa2016}, and \citet{marcote2016} in Fig.~\ref{fig:lc1}(b). The simulated light curves vary on scales of $\sim3-5$~d, consistent with all observations, and have highly correlated variability at 5.5 and 7.5~GHz, again consistent with the synchronized flux variation reported by \citetalias{keane2016}. Refractive scintillation can also explain the gentle spectral modulation across 2-18~GHz reported by \citet{Vedantham_2016}.

\section{The Origin of J0716-1900}\label{sec:4}
We have shown that refractive scattering can potentially explain the amplitude and timescale of the radio variations of J0716-1900. This explanation requires that the source is sufficiently compact, most plausibly a weak, radio-loud AGN. This explanation is consistent with preliminary results of VLBI observations \citep{bassa2016,marcote2016}. Here, we briefly discuss expected observational properties for this scenario. 

The typical flux density of $\sim 0.1-0.3$~mJy of J0716-1900 \citepalias{keane2016} corresponds to the radio power of
\begin{equation}
P_\nu = 9.4 \times 10^{22} \, {\rm W\,Hz^{-1}} \times \left( \frac{F_\nu}{0.1\,{\rm mJy}} \right) \left( \frac{D_{\rm L}}{2808\,{\rm Mpc}} \right) ^2.
\end{equation}
This power is consistent with the typical nuclear radio power of the nearby low luminosity AGNs (LLAGN) and elliptical galaxies \citep[e.g.][]{doi2011} that are thought to be powered by a hot accretion flow \citep[e.g.][]{yuan2014} or a faint jet \citep[e.g.][]{falcke2004}. This radio power is also consistent with the low-power end of the known blazars \citep[][]{liuzzo2013,massaro2015}. 

To avoid quenching the refractive scintillation, the source angular size must subtend $\theta_{\rm src} \lesssim 0.1$~mas, providing a lower limit on its brightness temperature:
\begin{eqnarray*}
	T_{\rm b} &=& \frac{c^2}{2k_{\rm B} \nu^2} \frac{F_\nu}{\pi (\theta_{\rm src}/2) ^2 \ln 2}\\
	&\gtrsim& 8.4 \times 10^{8} \,{\rm K}\times \left( \frac{\nu}{\rm 5.5 \,GHz} \right)^{-2} \left( \frac{F_\nu}{\rm 0.1 \, mJy}\right)\left( \frac{\theta_{\rm src}}{\rm 0.1 \, mas}\right)^2.
\end{eqnarray*}
The lower limit of the brightness temperature is compatible with low-power radio galaxies and blazars \citep[e.g.][]{liuzzo2009,liuzzo2013,piner2014}. Thus, in addition to the source size, the radio power and the brightness temperature are also reasonable as nuclear emission from a LLAGN, low-powered blazar, or weak AGN.

We note that \citet{Vedantham_2016} have very recently measured a flat radio spectrum for J0716-1900, which is generally seen in blazars \citep[e.g.][]{massaro2014}. The flat radio spectrum suggests that the majority of the arcsecond-scale flux density originates in the optically-thick radio core emission, which could be sufficiently compact to be scintillating, and other observations have revealed the presence of compact jet structure in elliptical galaxies without any feature of AGN in the optical/infrared spectrum \citep[e.g.][]{akiyama2016}. Thus, the radio spectrum supports the scenario in which the variability J0716-1900 is predominantly from interstellar scintillation.

\citet{williams2016} were the first to argue that the radio emission from J0716-1900 arises from an AGN and have noted that the observed variability is incompatible with standard afterglow evolution.　 Based on the deep VLA imaging study of \citet{Fomalont_1991} at 5~GHz, \citet{williams2016} further note that ${\sim}16$ sources above $100~\mu{\rm Jy}$ are expected per Parkes beam. Moreover, \citet{Fomalont_1991} found that most sources between $60-1000~\mu{\rm Jy}$ were unresolved (${<}1.5''$). These estimates are then favorable for ascribing the variability of J0716-1900 to refractive scintillation. However, we caution that the variability of J0716-1900 should not be directly compared with variability reported in other surveys at these frequencies because its line of sight is close to the Galactic plane and so has significantly stronger scattering than the median Galactic values. Consequently, the transition to weak scattering, where the refractive modulation index $m$ peaks, is at higher frequencies for J0716-1900 than for higher-latitude sources.

\section{Summary}
In short, the fast variation of J0716-1900 can be reasonably explained as refractive scintillation in the ISM and may not represent an afterglow associated with FRB150418. Both the analytical theory of refractive scattering and our numerical simulations show that the expected scattering of J0716-1900 can reproduce the observed timescales and modulation index at 5.5 and 7.5~GHz. They also naturally explain the synchronized modulation at these frequencies and the gentle modulation across the wider radio spectrum reported by \citet{Vedantham_2016}. Refractive scintillation requires that the source size is smaller than ${\lesssim}0.1$~mas (${\lesssim} 0.6$~pc at the source), which is consistent with preliminary results of VLBI observations \citep{bassa2016,marcote2016}. This source size corresponds to a brightness temperature $T_{\rm b} > 10^9~{\rm K}$, compatible with LLAGN and faint blazars. Our results would also apply to more compact emission, such as an FRB afterglow, and demonstrate that fast variability does \emph{not} necessitate a high Doppler factor. 

Our estimates of the scattering are not sensitive to assumptions about the location of the scattering screen, but our derived timescales are uncertain by a factor of several, both from the unknown velocity of the scattering material and the unknown distance of the scattering from the Earth. Also, our estimates of the scattering of J0716-1900 (from the NE2001 model) are uncertain by a factor of several. Finally, refractive scintillation can only cause the observed flux variability if the majority of the source flux originates in the compact core emission. The core dominance of J0716-1900 is therefore one of the key questions for continued studies and can be confirmed using VLBI. Even with these cautions and remaining uncertainties, it is evident that refractive scintillation is of fundamental importance for the interpretation of J0716-1900 and for radio identification of FRB afterglows more generally.

\acknowledgments
K.A.\ is financially supported by JSPS Postdoctoral Fellowships for Research Abroad and grants from the National Science Foundation (NSF). M.D.J.\ thanks the Gordon and Betty Moore Foundation for financial support (\#GBMF-3561). We also thank Dr.~Hiroshi Nagai for helpful discussions on FRB150418. We thank Dr.~Katherine Rosenfeld for writing ScatterBrane, the software used for our scattering simulations (\url{http://krosenfeld.github.io/scatterbrane}). 


\end{document}